\documentclass[twocolumn,showpacs,preprintnumbers,amsmath,amssymb,prb,fleqn]{revtex4}

\usepackage{epsfig}
\usepackage{dcolumn}
\usepackage{bm}
\usepackage{amsmath}

\begin{document}

\title{Modification of magnetic and transport properties of manganite layers in Au/La$_{0.67}$Sr$_{0.33}$MnO$_3$/SrTiO$_3$ interfaces}

\author{A. A. Sidorenko}
\email{sidorenko@fis.unipr.it}
\author{G. Allodi}
\author{R. De Renzi}
\affiliation{Dipartimento di Fisica e Unit\`a CNISM di Parma, Universit\`a degli Studi di Parma, Viale delle Scienze, 7A, 43100 Parma, Italy}
\author{S. Brivio}
\author{M. Cantoni}
\author{M. Finazzi}
\author{R. Bertacco}
\affiliation{L-NESS- Dipartimento di Fisica, Politecnico di Milano, Via Anzani 42, 22100 Como, Italy}
\date{\today}

\begin{abstract}
The effect of gold capping on magnetic and transport properties of optimally doped manganite thin films is studied. An extraordinary suppression of conductivity and magnetic properties occurs in epitaxial (001) La$_{0.67}$Sr$_{0.33}$MnO$_3$ (LSMO) films grown on SrTiO$_3$ upon deposition of 2 nm of Au: in the case of ultrathin films of LSMO (4 nm thick) the resistivity increases by four orders of magnitude while the Curie temperature decreases by 180 K. Zero-field $^{55}$Mn nuclear magnetic resonance reveals a significant reduction of ferromagnetic double-exchange mechanism in manganite films upon the gold capping. We find evidence for the formation of a 1.9-nm thick magnetic ``dead-layer'' at the Au/LSMO interface, associated with the creation of interfacial non double-exchange insulating phases.

\end{abstract}


\maketitle

\section{Introduction}

Ferromagnetic metallic oxide La$_{0.67}$Sr$_{0.33}$MnO$_3$ (LSMO) is a system with very complex and rich physics, whose electronic and magnetic properties may be altered by applying various perturbations such as electric and magnetic fields, strain, light, etc. High degree of spin polarization and colossal magnetoresistance make manganese perovskites, in particular the optimally doped ferromagnetic manganite LSMO, promising candidate for use in spintronic devices.\cite{Prellier2001,Renard2004,Viret1997} Such devices are fabricated as heterostructures by contacting thin manganite LSMO layer with another material, e.g., superconductors, insulators, organic materials, and normal metals.\cite{Bowen2003,Vasko1997,PhysRevLett.82.4288,Fratila2005,Dediu2002}
In addition, when manganites are embedded in heterostructures with other materials, new issues of physics, chemistry, and material science arise from the interfacial interaction.\cite{Ahn2006} Moreover, properties of thin LSMO layers are significantly different from those of the bulk and it is well established now, for instance, that strain strongly affects the bandwidth of thin LSMO layers depressing their metallic properties and giving rise to the phase separation and so-called dead-layer at the interfaces between LSMO films and substrates.\cite{Sidorenko2006} In some cases the suppression of the double exchange mechanism due to the orbital reconstruction is evident.\cite{Tebano2008} Furthermore, it is reasonable to assume that for such heterostructures the electronic and magnetic properties of LSMO layers at the interface are strongly affected by any conductive overlayer. Their interface with a metal is known to affect the chemical and electronic environments of manganites changing basic electronic parameters such as the Hubbard energy, the electronic bandwidth, or the exchange energies.\cite{Altieri2002}  This suggests that thin manganite films might be particularly suitable for devices where transport and magnetic properties could be tuned by the externally applied electric field.

Recently it was shown that the deposition of Au on top of 4-nm thick LSMO film produces a dramatic reduction of the Curie temperature T$_C $ ($\sim$185 K with respect to uncoated LSMO films of the same thickness) and reduces the value of the saturation magnetization.\cite{Bertacco2007, Brivio2007} A sizable T$_C $ reduction ($\sim$60 K) was observed even when an inert SrTiO$_3$ (STO) layer was inserted between the gold film and the 4-nm thick manganite layer, suggesting that this effect might have an electrostatic origin. A depletion of charge at the Ag/STO/La$_{0.67}$Ca$_{0.33}$MnO$_3$ interfaces and a contribution from phase separation in the manganite were also demonstrated in Ref.~\onlinecite{Boikov2004}

To gain a deeper insight into the structure of the interfaces in Au/LSMO junctions, and to better understand the physics of the observed phenomena, we carried out a more detailed investigation of  these heterostructures including the characterization of transport properties of the manganite films. According to the strict connection between transport and magnetism in manganites we found an extraordinary suppression of the conductivity accompanied to the suppression of magnetism. Magnetization and zero-field $^{55}$Mn NMR measurements have been performed to see if the changes in magnetic properties of the manganite thin layer upon Au capping are correlated with changes in the local atomic environment. We found a strong depression of the double exchange mechanism which explains the suppression of both transport and magnetic properties.

\section{Experimental details}

LSMO single crystal films with thickness ranging from 4 to 12 nm were grown by Pulsed Layer Deposition on SrTiO$_3$ substrates and then capped with a 1-2 nm thick gold layer by molecular beam epitaxy. AFM images taken on the LSMO 8-nm thick films, with and without Au capping layer, indicates atomically flat surfaces of the LSMO free layer with measured root-mean-square roughness about 0.2 nm The deposition of the Au film did not compromise the surface quality, within the limits of sensitivity of the AFM operated with standard tips. By repeating image acquisitions in different parts of the sample, it was verified that the samples display a very small amount of droplets, which were estimated to cover only 1\% of the whole sample surface.\cite{Ghidini}

In-plane magnetization was measured by using a commercial superconducting quantum interference device (SQUID), 5 T Quantum MPMS, in the field of 100 Oe. We avoided spurious contributions to the measured signal, other than from the film and the substrate, by using diamagnetic sample holders with uniform mass and magnetic moment distribution along the whole SQUID scanning length. The magnetization data of the manganite films were corrected for the observed diamagnetic contribution of the substrate and the top gold layers.

Resistance measurements as a function of temperature were performed with a standard four points probe apparatus without patterning the film, with probes disposed on top of the films/heterostructures. A Keithley 2601 was used as a current source injecting a constant current of 100 nA, and resistivity values have been extracted according to Van der Paw relations from the knowledge of the film thickness.

Zero-field $^{55}$Mn NMR spectra were collected with the home-built broadband fast-averaging NMR spectrometer HyReSpect \cite{Allodi2005} on a tuned probe circuit at T=1.6 K. The $^{55}$Mn NMR spectra were obtained by Fourier transforming two pulse sequence. The plotted NMR spectra are always corrected rescaling their amplitudes by $\omega^{2}$ and enhancement factor $\eta$.\cite{Turov1972}

\section{Results and Discussion}

\begin{figure}
\centerline{\psfig{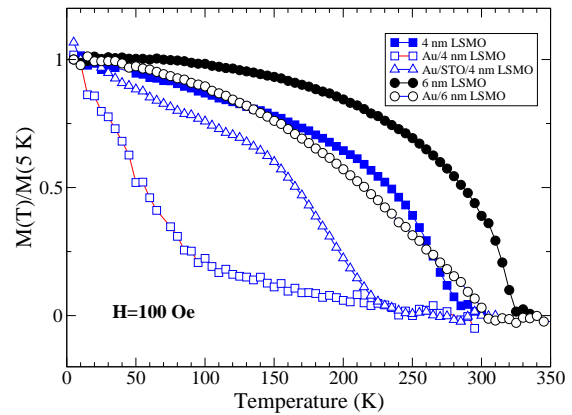}}
\caption{(color online) Normalized magnetization vs. temperature measured at 100 Oe in variant heterostructures. The thickness both of the gold layer and of the STO buffer layer is 2 nm.}
\label{fig:Tc_heterostructures}
\end{figure}

The temperature dependence of the reduced magnetization measured in different heterostructures at the magnetic field of 100 Oe is shown in Fig.~\ref{fig:Tc_heterostructures}. The Curie temperature T$_C$ obtained as the temperature corresponding to the higher temperature inflection point of M(T) decreases with the film thickness as expected for thin magnetic films: $325\pm5$ K for 6 nm thick LSMO film (closed circles) and $280\pm5$ K for 4 nm thick LSMO film (closed squares). Very significant lowering ($\sim185\pm11$ K) of the T$_C$ is found after depositing a 2 nm thick gold layer on top of  the 4 nm LSMO film (see open squares in Fig.~\ref{fig:Tc_heterostructures}) while for the 6 nm thick LSMO film the T$_C$ reduction is much smaller ($\sim10\pm3$ K) but is still clearly seen (open circles). The effect of the gold layer deposition disappears completely when the thickness of the manganite layer exceeds 8 nm. This indicates that the phenomenon is limited to an interfacial layer, so that it is no more visible in a bulk measurements like SQUID when the thickness of the film exceeds by far that of the interfacial layer affected by the proximity of the Au layer.

Many different processes could be involved in altering the magnetic properties of LSMO after the deposition of an overlayer of Au. First of all, it is well known that the magnetic properties, in particular the Curie temperature of manganites are extremely sensitive to oxygen content.\cite{Renard1999,Trukhanov2004,Abdel2000,Wad2002} Several previous experiments indicated the creation of an oxygen depleted layer of LSMO interfaces in contact with such metals as Nb, Al, Pd, and Au.\cite{Mieville1998, Stadler2000,Plecenik2002} Therefore one might suppose the formation  of a La$_{0.67}$Sr$_{0.33}$MnO$_{3-\delta}$ ($\delta$ - oxygen vacancies) layer caused by the out-diffusion of oxygen from the manganite surface to the gold layer. 
However, detailed XPS measurements did not reveal any out-diffusion of oxygen from the manganites film nor any strong interdiffusion at the interface.\cite{Bertacco2007, Brivio2007,Petti2008} 

In addition, inserting, e.g., a chemically inert STO buffer layer between the gold layer and manganite film to create a Au/STO/LSMO heterostructure we can avoid the possible chemical reaction at the interface and any out-diffusion from the manganite film to the capping Au layer. Blue triangles in Fig.~\ref{fig:Tc_heterostructures} display the temperature dependence of magnetization of a  Au(2nm)/STO(2nm)/LSMO(4nm) heterostructure  measured in a magnetic field of 100 Oe. In spite of the introduction of such STO buffer layer a sizable decrease of the T$_C$ ($\Delta T_C\sim 60$ K) is still found. It should be noted that the magnetization measurements of similar STO/LSMO/STO heterostructures, without gold, did not reveal any effect of thin STO buffer layer on the magnetic properties of LSMO layer.\cite{Bertacco2007} 

\begin{figure}
\centerline{\psfig{file=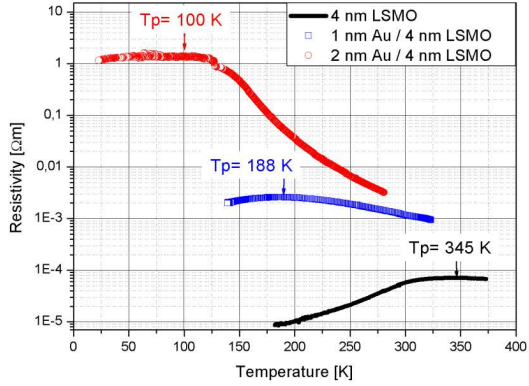}}
\caption{(color online) Resistivity as a function of temperature for 4 nm LSMO (solid line), 1-nm Au/4-nm LSMO (squares) and 2-nm Au/4nm LSMO (circles).}   
\label{fig:resistivity}
\end{figure}

The other effect that could be responsible for the reduction of T$_C$ in LSMO is the strain induced by the gold overlayer.\cite{Millis2000, Kanki2001} A biaxial strain increases the energy differences between $e_g$ levels imposed by Jahn-Teller interaction which reinforces the electrons tendency to become more localized thus producing a decrease of T$_C$.\cite{Millis2000} However, in our case the strain effects can also be reasonably ruled out since the Au film is thin (2 nm) and not epitaxial. Furthermore, in the case of heterostructures involving the STO buffer layer, any strain would be partially released by the STO spacer. The influence of a well-defined reversible biaxial strain on the magnetization and T$_C$ of epitaxial LSMO films was investigated in Ref.~\onlinecite{Thiele2007}. By extrapolating the linear dependence of $T_C$ versus the in-plane strain to the (unreasonably high) value corresponding to the Au-STO lattice mismatch ($\simeq4\%$), one can estimate a maximum $T_C$ reduction of less than 40 K. Therefore, strain cannot explain the observed $T_C$ decrease, at least for 4 nm thick LSMO film.

In Fig.~\ref{fig:resistivity} the resistivity curves for a 4 nm thick LSMO films (solid line) and the heterostructures 1-nm Au/4-nm LSMO (squares) and 2-nm Au/4-nm LSMO (circles) are shown. The 4 nm thick LSMO film without gold displays a metal-insulator transition temperature (T$_P$) and resistivity values typical of films of this thickness.\cite{Brivio2007} A gold capping layer of 1 nm produces a strong increase of the peak resistivity (more than one order of magnitude) and a shift of the metal-insulator transition to lower temperatures (by 160 K). Note that the gold overlayer gives a negligible contribution to the electrical conduction as expected for films with a thickness comparable with the mean free path of the conduction electrons. Furthermore, the gold electric properties could also suffer from its non-epitaxial growth which leads to a polycrystalline and possibly porous layer. With a 2 nm capping layer the effects are even larger: the critical temperature is further reduced (down to 100 K) and the peak resistivity value is raised by about four orders of magnitude with respect to the free LSMO film. 
T$_P$ is distinct from T$_C$, as observed also in thicker samples,\cite{Bertacco2005} but they correlated with each other and they both reduce drastically with Au capping. The progressive increase of resistivity with nominal Au thickness suggests the formation of a separate insulating phase at the interface, in competition with the conductive phase, whose fine microstructure in the plane of the surface could also influence the measured macroscopic electrical resistance by controlling percolation. Phase separation could then be a key point in the explanation of the observed phenomena.

\begin{figure}
\centerline{\psfig{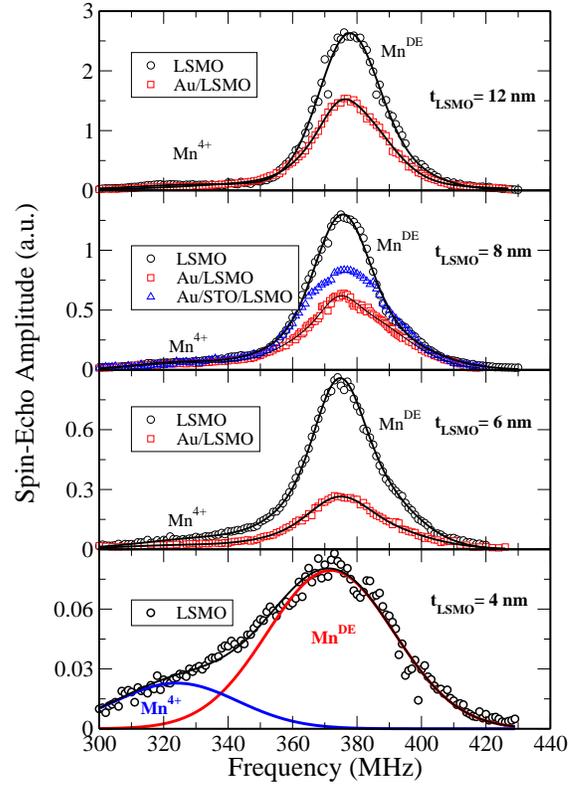}}
\caption{(color online) Zero-field $^{55}$Mn NMR in variant heterostructures: circles - LSMO/STO, squares - Au/LSMO/STO, and triangles - Au/STO/LSMO/STO. The thickness of both the STO buffer layer and Au layer is fixed to be 2 nm, while the thickness of the LSMO layer is t$_{LSMO}$= 4, 6, 8, 12 nm. Solid lines demonstrate the best fit to the experimental data.}
\label{fig:NMR_spec}
\end{figure}

To determine the microstructural origin of changes in the magnetic behavior of the LSMO film that resulted from the addition of the gold layer we have carried out zero-field $^{55}$Mn NMR experiments. It is very well established that in the mixed valence manganites different manganese states yield distinct contributions to the NMR spectra. The localized Mn$^{4+}$ state gives rise to a peak between 310 and 330 MHz, whereas localized Mn$^{3+}$ resonance varies between 350 MHz and 430 MHz. The signal corresponding to the Mn$^{DE}$ state from the mixed valence metallic region is associated with a fast hopping of electrons among Mn sites and it shows up as a relatively narrow peak at an intermediate frequency in the 370-400 MHz range.\cite{Matsu, Allodi1, Kapusta, Savosta2}

\begin{figure}
\centerline{\psfig{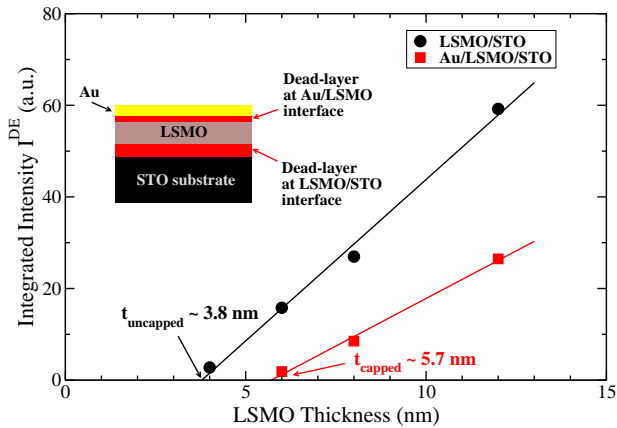}}
\caption{(color online) Thickness dependence of the NMR integrated intensity I$^{DE}$ measured in the samples with (squares) and without (circles) gold overlayer. Solid lines are linear fit to the experimental data. The inset shows schematically the formation of the dead-layers at Au/LSMO/STO interfaces.}
\label{fig:NMR_amp}
\end{figure}

Fig.~\ref{fig:NMR_spec} shows the $^{55}$Mn NMR spectra (solid lines represent a best fit) that we collected from several heterostructures with and without the gold overlayer. In all spectra at least two distinct lines at $f^{4+}\approx322$ MHz and $f^{DE}\approx376$ MHz are observed and may be attributed to two different phases with the localized charges (Mn$^{4+}$ state) and with itinerant carriers (Mn$^{DE}$ state), respectively.\cite{Bibes2001,Sidorenko2006} As can be seen the intensity of $f^{DE}$ peak (the area under curve) scales with the film thickness (see also Fig.~\ref{fig:NMR_amp}), and it is expected to vanish for a LSMO thickness corresponding to the dead-layer, whereas the intensity of the peak at $f^{4+}$ is practically unvaried.\cite{Sidorenko2006}

The spectra of the LSMO films terminated with Au cappings also reveal two resonance lines approximately at the same frequencies $f^{4+}$ and $f^{DE}$ (red squares in Fig.~\ref{fig:NMR_spec}). 
Since the intensity of the NMR line is proportional to the number of the nuclei in the specific electronic and magnetic state, the change of the Mn$^{4+}$/Mn$^{DE}$ ratio due to the presence of the gold layer can be measured by NMR directly. As can be seen in Fig.~\ref{fig:NMR_spec} a marked reduction of the $f^{DE}$ signal is found in the gold terminated samples, indicating that the capping layer has a direct influence on this specific electronic state (double exchange state) of the manganite film. The spectroscopic sensitivity of NMR allows us to detect quite a large effect already on the thickest 12 nm structures. The reduction is  larger for thinner films, as it would be expected in the case Au induces a constant finite thickness depletion layer in LSMO. The introduction of a thin STO spacer between Au and LSMO layers results in a partial recovery of the Mn$^{DE}$ peak intensity corresponding to an increase of the amount of ions in the Mn$^{DE}$ state (in Fig.~\ref{fig:NMR_spec} the effect of the STO buffer layer is shown only for the 8-nm thick LSMO layer). However, even in such a case the effect of the gold capping layer on the manganese double-exchange state in the manganite layers is still observed. For the case of the 4 nm thick LSMO film with the gold capping the zero-field NMR signal is beyond the sensitivity of our NMR spectrometer and was not detected revealing that the majority of the LSMO film is in a non-DE state. Notice that it does not conflict with the SQUID results, that show the presence of a low temperature magnetization even for 4 nm of LSMO capped with gold, since manganites may give rise to a variety of insulating ferromagnetic components, such as spin-canted phases, superexchange ferromagnets or blocked superparamagnetic clusters. They are not DE phases and their contribution to NMR is negligible while they are still visible in SQUID.

The integrated intensity of Mn$^{DE}$ state, obtained from the fit of the NMR spectra, versus LSMO film thickness for the samples with (red squares) and without (black circles) the gold top layer is plotted in Fig.~\ref{fig:NMR_amp}. From the linear fit (solid lines) of the experimental points we can find that the critical thickness when the $f^{DE}$ signal disappears is $t_{uncapped}\approx3.8$ nm and $t_{capped}\approx5.7$ nm for gold uncapped and capped LSMO films, respectively.  We should note that this critical thickness $t_{uncapped}$ in uncapped LSMO/STO roughly coincides with the thickness ($\sim$3 nm) of the dead-layer below which the manganite films appear to be insulating.\cite{Bertacco2007} The higher critical thickness in the samples with the gold capping indicates an additional reduction of the amount of the ferromagnetic Mn$^{DE}$ ions at Au/LSMO interface (see inset in Fig.~\ref{fig:NMR_amp}). A simple estimation gives that the thickness of such a dead-layer at the interface between gold and LSMO layers is $t_{Au/LSMO}\approx1.9$ nm.

We would like to stress that if an oxygen deficiency took place in Au/LSMO heterostructures, the positions of the resonance lines would be different. An oxygen deficiency decreases the Mn$^{4+}$ concentration and affects the crystal lattice, reducing the transfer interaction of $e_g$ electrons and weakening the ferromagnetic double exchange interactions,\cite{Trukhanov2004,Abdel2000} and, hence, is expected to have an influence on the hyperfine interactions of manganese ions, which can be measured by NMR.\cite{Guevara1997} The expected hole concentration (Mn$^{4+}$) is given by simple relation $c=0.33-2 \cdot \delta$ so assuming the oxygen vacancies $\delta = 0.05$ (it corresponds to $\Delta T_c\approx60$ K in the case of LSMO)\cite{Abdel2000} we find that $c$ is of the order of 0.2. However, such hole concentration would give rise to entirely different zero-field $^{55}$Mn NMR spectra.\cite{Allodi1997,Papavassiliou}

\section{Conclusions}

To conclude, we studied the effect of the gold capping on magnetic and transport properties of ultrathin manganite films in Au/LSMO/STO heterostructures. A very large ($\sim185\pm11$ K) reduction of the Curie temperature is found after depositing a 2-nm thick gold layer on top of the 4-nm thick LSMO film. Resistivity measurements show a lowering of the metal-insulator transition temperature  and an extraordinary increase of the peak value of resistivity (by four orders of magnitude) which is consistent with the parallel suppression of the magnetic character. This underlines the strict correlation between magnetism and transport arising from the double exchange mechanism. The decrease of T$_C$ is evident even after inserting the chemically inert STO buffer layer between the gold layer and LSMO film. Zero-field $^{55}$Mn NMR reveals that the gold deposition suppresses the Mn$^{DE}$ state. Within the framework of double exchange theory this information is naturally reconciled with the transport evidence of an interfacial high resistance layer. Our experimental findings point to the fact that gold capping produces the formation of a dead layer, insulating and, hence, not of the double exchange type, at the interface. 

\section*{ACKNOWLEDGMENTS}

This work has been funded by Consorzio Nazionale Interuniversitario per le Scienze Fisiche della Materia (CNISM) as part of a Progetto d'Innesco della Ricerca Esplorativa 2005, and partly by Netlab NanoFaber and FPVI STREP OFSPIN. The authors wish to thank Dr. Massimo Ghidini for fruitful discussions.

\bibliography{Text}
\end{document}